\documentclass{iopart}
\usepackage{iopams}  
\usepackage{graphicx}
\begin{document}

\rapid[Electrostatic considerations affecting the calculated HOMO-LUMO gaps]{Electrostatic considerations affecting the calculated HOMO-LUMO gap in protein molecules}

\author{Greg Lever$^1$, Daniel J Cole$^{1,2}$, Nicholas D M Hine$^3$, Peter D Haynes$^3$ and Mike C Payne$^1$}

\address{$^1$ Theory of Condensed Matter group, Cavendish Laboratory, 19 JJ Thomson Ave, Cambridge CB3 0HE, United Kingdom}
\address{$^2$ Department of Chemistry, Yale University, New Haven, Connecticut 06520-8107, United States}
\address{$^3$ Departments of Materials and Physics, Imperial College London, Exhibition Road, London SW7 2AZ, United Kingdom}

\ead{gl319@cam.ac.uk}

\begin{abstract}
A detailed study of energy differences between the highest occupied and lowest unoccupied molecular orbitals (HOMO-LUMO gaps) in protein systems and water clusters is presented. 
Recent work questioning the applicability of Kohn-Sham density-functional theory to proteins and large water clusters (E. Rudberg, 
\emph{J. Phys.: Condens. Mat.} 2012, \textbf{24}, 072202) has demonstrated vanishing HOMO-LUMO gaps for these systems, which is generally attributed to the treatment of exchange in the functional used. 
The present work shows that the vanishing gap is, in fact, an electrostatic artefact of the method used to prepare the system. 
Practical solutions for ensuring the gap is maintained when the system size is increased are demonstrated. 
This work has important implications for the use of large-scale density-functional theory in biomolecular systems, particularly in the simulation of photoemission, optical absorption and electronic transport, all of which depend critically on 
differences between energies of molecular orbitals.  
\end{abstract}

\section{Introduction}

Density-functional theory (DFT)~\cite{Kohn,Sham} simulations are becoming increasingly
widespread for simulating biological systems at the level of
individual atoms and electrons.
The advantages of DFT over, for example, molecular mechanics (MM)
force fields include correct treatment of long-range polarisation,
charge transfer, bond breaking and forming, electronic states and, by association, spectroscopy. 
In addition, transition metals,
unusual ligands or functional groups that are difficult to correctly parameterise within force fields are also accurately treated.
A limiting factor in applying conventional first-principles
approaches to large systems 
is the unfavourable computational 
requirements which typically increase as the third (or greater) power of the number of atoms in
the system.
However, methods to overcome this bottleneck using linear-scaling approaches to DFT have been under development for over 
a decade~\cite{Goedecker_Rev} and are still advancing today~\cite{Bowler_2012}, allowing system sizes on the order of tens of 
thousands of atoms to be 
routinely accessed~\cite{onetep_thousands}. 

As larger systems become accessible, the application of density-functional methods to 
systems of biological interest is becoming
increasingly common.
The FHI-aims package~\cite{FHI-aims} has been applied to folding
processes in helices and polypeptides~\cite{Tkatchenko_2011}.
The TeraChem package~\cite{Terachem} has
recently been used to optimise the structures of more than 50 polypeptides of sizes ranging up to 590 atoms,
predicting protein structure with the same accuracy as empirical force fields paramaterised extensively for this 
purpose~\cite{Kulik_2012}.
Linear-scaling DFT calculations have been reported on a dry
DNA model comprising 715 atoms using the SIESTA code~\cite{Siesta_DNA} and a B-DNA decamer  
using the CONQUEST code~\cite{Conquest} with explicit water molecules and counter ions~\cite{Conquest_DNA}.
The \textsc{ONETEP} code~\cite{onetep_1}, applied in the present
investigation, has previously been used to perform geometry
optimisations to characterise binding energetics of small
molecules to the metalloprotein myoglobin~\cite{Cole_2012} and to
measure binding of small molecules to T4 lysozyme in
solution~\cite{Dziedzic_2012}.
Recent developments in the simulation of optical spectroscopy~\cite{Ratcliff_2011},
dynamical mean field theory with applications to human
respiration~\cite{Weber_2012_arxiv} and methods to aid interpretation
of the electronic structure~\cite{Lee_12, Lee_sub} 
broaden the scope of biomolecular simulations still further.

Despite considerable interest in the use of {\it ab initio} simulations
for the study of complex biomolecular systems, there is a growing concern that
DFT, in conjunction with pure exchange-correlation functionals (those defined as containing no Hartree-Fock exchange), may be
inappropriate for the simulation of large molecular clusters.
In particular, it has recently been shown that there are unphysical vanishing
HOMO-LUMO gaps in systems such as proteins~\cite{Rudberg_2012} and
even water clusters~\cite{Rudberg_2011}. This can lead to poor convergence during
the self-consistent electronic structure optimisation procedure.
Poor self-consistent field (SCF) convergence has been found for BLYP and B3LYP DFT functionals~\cite{Isborn_2011} and for 
LDA and PBE functionals when simulating large glutamic acid-alanine helices~\cite{Ergo}.
Attempts to optimise polypeptide structures \emph{in vacuo} using the 
TeraChem package with BLYP and B3LYP functionals reported a lack of SCF 
convergence for many of the peptides simulated~\cite{Kulik_2012}.
Similar problems have been observed when using molecular fractionation with conjugated caps to compute \emph{ab initio} 
binding energies in vacuum for protein-ligand complexes of between 1000 and 3000 atoms~\cite{Antony_2012}. 

SCF convergence issues are widely blamed upon the well-known
phenomenon of pure functionals under-estimating the HOMO-LUMO 
gap~\cite{Perdew_81, Godby_88, Seidl_96}.
However, while the lack of the derivative discontinuity of the
exchange-correlation potential at integer particle numbers and errors
in the single-particle eigenvalues resulting from the approximate
nature of the functional itself will indeed reduce the gap, there is
no obvious reason why the effect should worsen at larger system
sizes~\cite{Godby_88}.
Indeed, it has been shown that recovery of a sizeable gap and consequently robust
self-consistent convergence of the electronic energy levels is
possible by including electrostatically embedded point charges to represent water molecules around
an inner cluster treated with quantum mechanics (QM)~\cite{Rudberg_2012,doCouto_05} or by simulating the 
system in a dielectric medium~\cite{Antony_2012}.
These results point to the possibility that the vanishing gap is a
surface effect and not an inherent difficulty with pure Kohn-Sham DFT.

In this communication, we use the linear-scaling DFT code ONETEP~\cite{onetep_1} to investigate the 
HOMO-LUMO gap of water clusters and protein systems. As in previous studies~\cite{Rudberg_2012} we find that the 
gap often vanishes \emph{in vacuo}.
However, we provide conclusive evidence that the issue is not related
to the use of a pure exchange-correlation functional but rather is
a result of the approach used to prepare the system.
We suggest a number of practical measures for preparing large systems
that do not lead to a vanishing HOMO-LUMO gap and thus open the way for continued
investigations of biomolecular systems with Kohn-Sham DFT.
%
\section{Computational Method}
The present work uses ONETEP~\cite{onetep_1}, a linear-scaling DFT package designed for use on parallel 
computers~\cite{onetep_implementation}
that combines near-complete basis set accuracy with a computational cost that scales linearly with the number of atoms. 
This allows an accurate QM description of systems of thousands of atoms~\cite{onetep_thousands}, including entire 
proteins~\cite{Cole_2012,onetep_interactions, onetep_proteins}. 
The ONETEP formalism is based on a reformulation of conventional 
Kohn-Sham DFT~\cite{Kohn,Sham}
in terms of the single-particle density matrix
\begin{equation}
\rho(\mathbf{r},\mathbf{r^\prime})=\sum_{\alpha\beta}\phi_\alpha(\mathbf{r})K^{\alpha\beta}\phi_\beta^\ast(\mathbf{r^\prime})
\end{equation}
where $\phi_\alpha(\mathbf{r})$ are non-orthogonal generalised Wannier functions (NGWFs)~\cite{onetep_2} that 
are localized in real space. 
The density kernel $K^{\alpha\beta}$ is a representation of the density matrix in the biorthogonal duals of the NGWFs. 
ONETEP achieves linear scaling by exploiting the 
``nearsightedness'' of the single-particle density matrix in non-metallic systems~\cite{Kohn-nearsighted-1996, Kohn-nearsighted-2005}.
In practice, linear scaling arises from enforcing strict localisation of the NGWFs onto atomic regions and through the optimisation of the density kernel 
and NGWFs, subject to localisation constraints. 
Optimising the NGWFs \emph{in situ} allows for a minimal number of atom-centred orbitals to be used whilst 
maintaining plane-wave accuracy. 
The basis set underlying the NGWFs consists of periodic cardinal sine (psinc) functions~\cite{onetep_4}.
The use of a plane-wave basis allows an unbiased approach to 
DFT calculations with systematically improvable accuracy by varying a single parameter similar to the energy cutoff in 
conventional plane-wave DFT packages.
QM calculations have been performed in ONETEP using the PBE gradient corrected 
exchange-correlation functional~\cite{PBE-func}. 
ONETEP will converge to a minimised electronic energy irrespective of the filling 
of the states that is applied at the beginning of a simulation. No density kernel truncation has been performed 
in the current calculations.  
The kernel is optimised using a combination of methods~\cite{Haynes_08}, which has the effect 
of ensuring integer occupancies of zero or one for all single-electron eigenstates of the Hamiltonian, which is 
appropriate in bulk insulators with a clearly defined band gap or, as is the case in the systems in the present work, in
 isolated systems with a clear HOMO-LUMO gap.
In cases where the systems have a zero gap, traditional linear-scaling algorithms for the optimisation of the density
kernel are not suitable and often converge to unphysical solutions not corresponding to integer occupation 
of the valence eigenstates.
In such cases we have indicated that the system did not converge. 

Where indicated in the text the use of implicit solvent has been made. 
The implicit solvent model is fully self-consistent and based on direct solution of the inhomogeneous Poisson equation. 
The approach involves defining the solute cavity in terms of an isosurface of the electron density~\cite{Fattebert_Gygi} and has been
 implemented in ONETEP~\cite{Dziedzic_2012,Dziedzic_2011}. 
 In addition, the use of electrostatic point charges (denoted by QM/EE in the text) has been made. 
 Electrostatic embedding can significantly reduce the computational cost associated with density-functional calculations 
 by representing a selected portion of the total system in terms of highly localised classical charge distributions that are electrostatically 
 coupled with the quantum system, representing the effect of the environment in which the quantum 
 system is embedded~\cite{onetep_embedding}.
ONETEP parameters are described in detail in the Supplementary Methods.
%
\section{Water Clusters}\label{sec:WaterClusters}
\begin{figure}[h]
\centering
\includegraphics[scale=0.3]{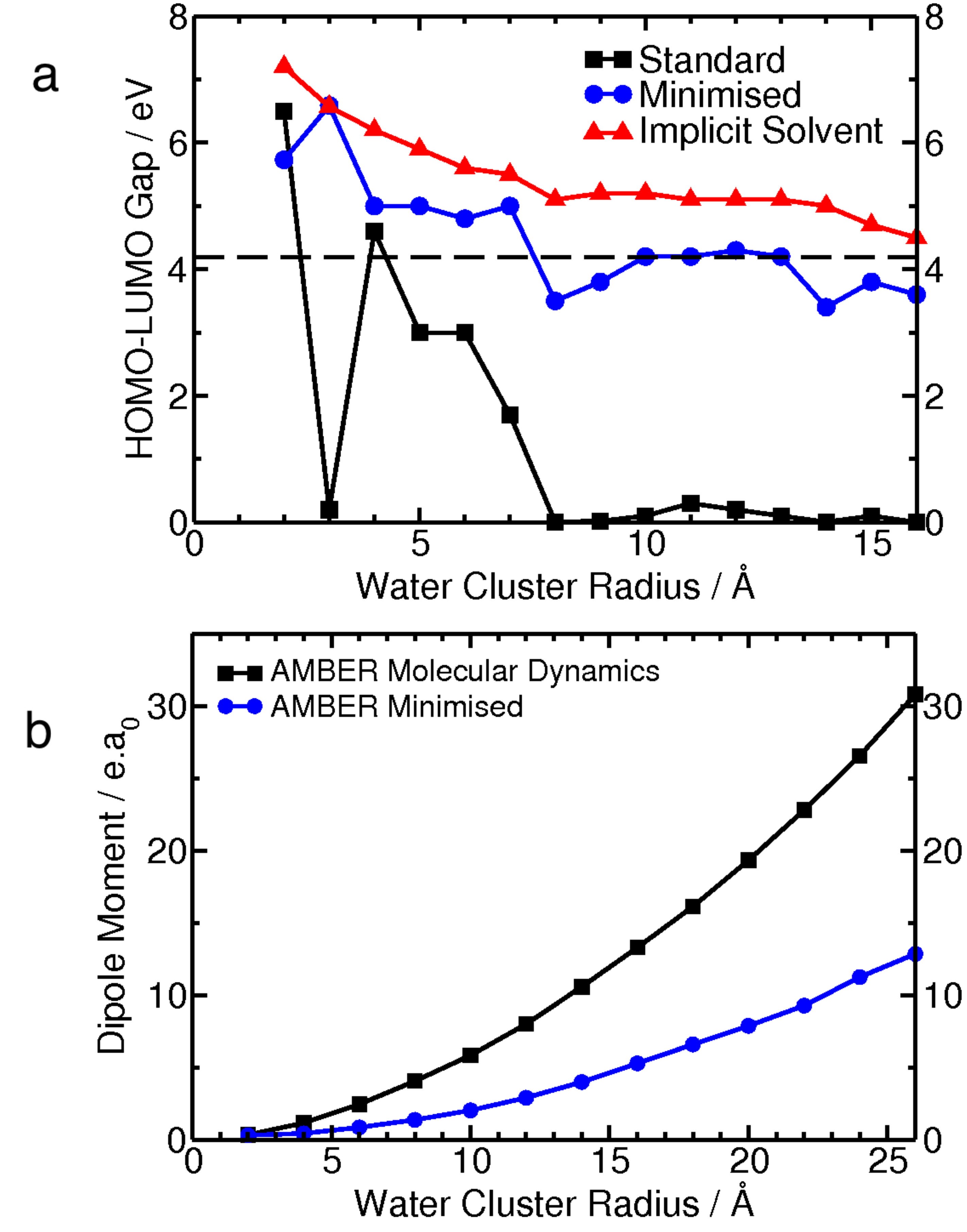}
\caption{(a) DFT HOMO-LUMO gaps of water clusters of increasing radius
  extracted from a larger 50~\AA{} cube of water equilibrated at 300~K
  using classical MD. Black line: Extracted straight from bulk
  water. Blue line: After classical minimisation is performed on
  each extracted cluster. Red line: Simulating the extracted clusters
  in implicit solvent.  Dashed line: HOMO-LUMO gap of bulk
  water. (b) Average dipole moments of water clusters of increasing
  radius calculated using the TIP3P point charge model. Black line:
  averaged over 5000 snapshots extracted from a larger 50~\AA{} cube
  of water. Blue line: averaged over 1800 snapshots extracted from the
  bulk and minimised using an MM force field.}
\label{fig:gap+dip_plot}
\end{figure}

The correct treatment of water is vital for realistic simulations of
biomolecular environments.
However, recent large-scale DFT simulations with pure functionals have
encountered SCF convergence problems when simulating isolated water
clusters, due to the HOMO-LUMO gap decreasing to zero when the cluster
radius becomes larger than approximately 10~\AA{}~\cite{Rudberg_2012}.
We begin by using the linear-scaling DFT code ONETEP to calculate the
band gap of a 2010-atom periodic supercell of bulk water,
equilibrated using the classical molecular dynamics package
AMBER~\cite{amber11} (Supporting Methods).
Consistent with previous DFT calculations of water employing the PBE
functional and as expected for a bulk insulator, the system has a
clearly defined band gap of 4.2~eV.
We have also calculated the HOMO-LUMO gap of spherical water clusters
of increasing radius extracted from a larger 50~\AA{} cube of water
equilibrated at 300~K with classical molecular dynamics (MD).
For small clusters such as these, quantum confinement would generally
be expected to result in a HOMO-LUMO gap greater than that of the
bulk, although one might expect the value of the cluster gap to tend
to the bulk band gap with increasing system size.
However, in agreement with previous work~\cite{Rudberg_2012},
Figure~\ref{fig:gap+dip_plot}(a) (black line) shows that the HOMO-LUMO
gap quickly approaches zero for systems of more than approximately 200
atoms.
This observation has therefore led many to question the applicability
of pure DFT functionals to large systems, such as water clusters and
proteins~\cite{Kulik_2012,Rudberg_2012,Antony_2012,doCouto_05,Grimme_2012}.

\begin{figure}[h]
\centering
\includegraphics[scale=0.12]{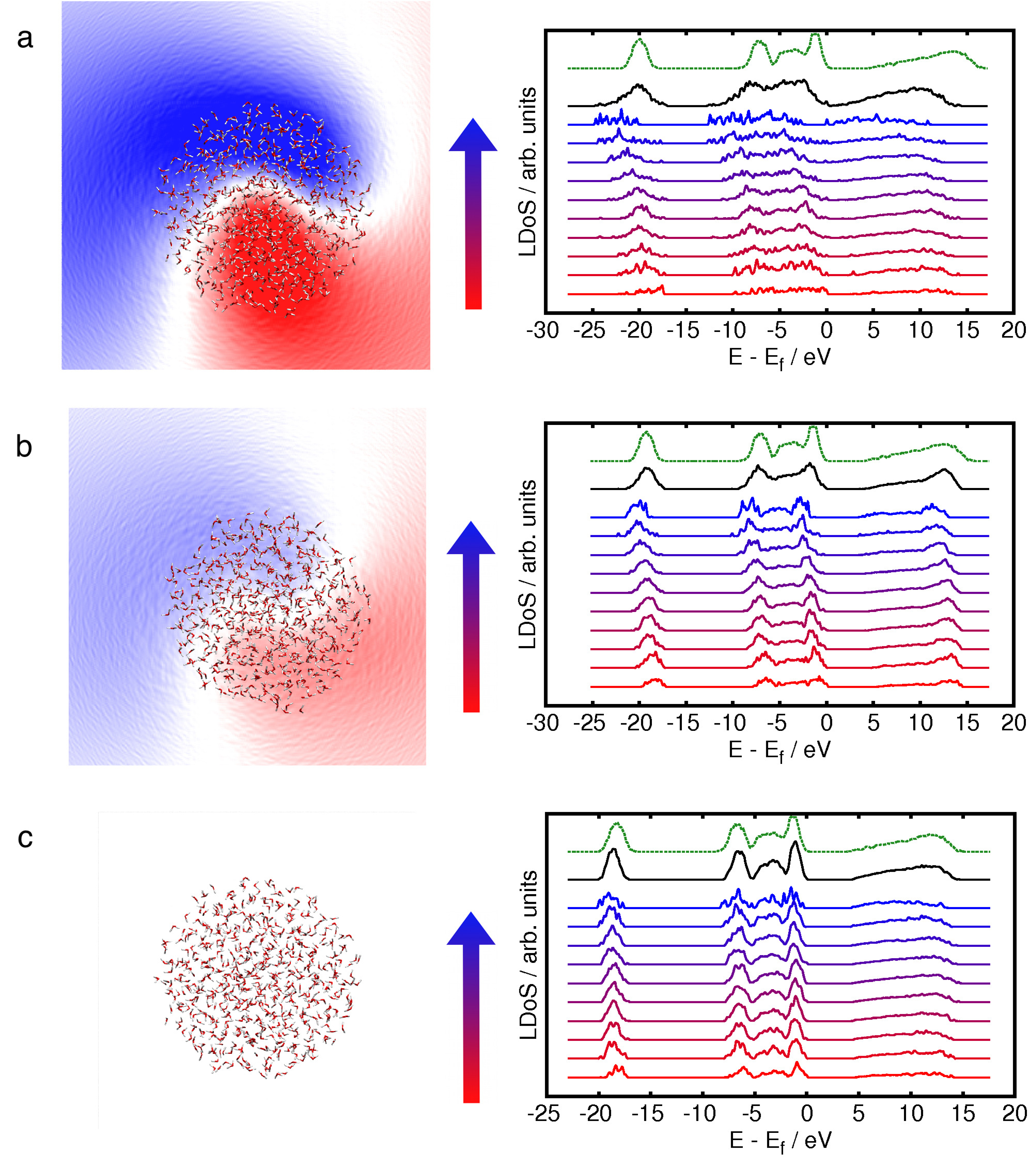}
\caption{Electrostatic potential for a 16~\AA~radius water cluster  
  and local density of electronic states for groups of atoms as a
  function of position along the dipole moment vector of the cluster.
  The dipole moment vector (coloured arrow) runs from the red line to
  blue. The black line is the total density of states and the green
  dashed line is the total density of states for bulk water.  Each
  line in the LDoS plot is normalised by the number of molecules
  contained in the slab. The electrostatic potential ranges from -0.3
  V (red) to +0.3 V (blue). The slice is 24.6~\AA~behind
  the water cluster. (a) Snapshot extracted from bulk water. The
  dipole moment is high, the LDoS is strongly dependent on position
  relative to the dipole moment vector, and the total range of states
  is much wider than for bulk water. (b) After classical minimisation
  of the same snapshot. (c) Simulated using the ONETEP implicit solvent model. In
  both cases, the dipole moment is reduced and the DoS closely
  resembles that of the bulk.}
\label{fig:LDoS_plot_16}
\end{figure}

However, our results showing the insulating nature of very large
supercells of bulk water indicate that the system size in itself is
not a problem but, rather, the small HOMO-LUMO gap is caused by the
vacuum-water interface.
Bulk water consists of a continuous hydrogen-bonded network of water
molecules, each of which has an intrinsic dipole moment.
The process of extracting a cluster, freezing the atomic positions and
surrounding with vacuum potentially results in a large surface dipole
being created.
A rough estimate illustrates this: the dipole moment of a single isolated water molecule $(0.73~e.a_0)$ produces a potential
difference of 0.4~V between opposing points on a sphere of radius 5~\AA{} with the molecule at its centre. 
While the molecular dipoles inside the cluster are oriented so as to mostly cancel any long-ranged effect,
those on the surface are not compensated by their neighbours, so a large cluster will retain a large net dipole moment.
Indeed, Figure~\ref{fig:gap+dip_plot}(b) (black line), which shows the average
dipole moment of clusters of water molecules calculated using a TIP3P
point charge model for water~\cite{TIP3P}, reveals that the dipole moment increases
with system size as the created surfaces become larger in area. 
A similar trend in dipole moment is observed for QM calculations of
single snapshots of water clusters of increasing radius (Figure S1)
and Figure~\ref{fig:LDoS_plot_16}(a) plots the DFT electrostatic
potential on a plane behind the 16~\AA{} water cluster, revealing a
clear dipolar potential.

We have demonstrated that water clusters extracted from the
equilibrated bulk display large multipole moments, as measured both by
MM and QM.
What effect does this have on the computed HOMO-LUMO gap?
Figure~\ref{fig:LDoS_plot_16}(a) also shows the local density of
electronic energy states (LDoS) for a 16~\AA{} water cluster.
To determine this quantity, clusters are nominally divided into 10
slabs perpendicular to the dipole moment, defined as the
$z$-direction, and the slab LDoS is defined as the sum of the
contributions to the total DoS from the local orbitals centred on the
atoms in each slab.
We observe a clear shift in the LDoS as a function of the
position along the dipole moment vector, with the electric field pushing some states
higher in energy and some lower. 
Analogous to the concept of Fermi-level pinning in polar semiconductor
nanorods, where the Fermi energy coincides with a finite density
of states at either end of the rods~\cite{Avraam_Fermi,Hine_nanorods},
the Fermi energy coincides with a non-zero density of states on opposite surfaces
of the water cluster.
We expect the HOMO-LUMO gap to disappear completely when the radius of
the water cluster increases to the point where the variation of the surface potential is of
sufficient magnitude to bridge the gap.

Given the apparent electrostatic origin of the vanishing gap, we
hypothesise that there is no fundamental problem in the use of pure
functionals in simulations of large systems, but simply that the issue
manifests itself at smaller system sizes than it would for hybrid
functionals which have an inherently larger gap.
We would, therefore, expect any method that corrects these surface
effects to also restore the HOMO-LUMO gap, which is consistent with
observations made in the literature.
It has been shown previously that the HOMO-LUMO gap may be restored by
embedding classical point charges outside the electron distribution to
represent, for example, the aqueous environment of the water or
protein cluster~\cite{Rudberg_2012}.
In these cases, no significant changes occurred to the electronic
density of the inner water molecules and the only significant changes
in electronic density were observed on water molecules close to the
surface~\cite{doCouto_05}.
Further, findings that a dielectric medium with a relative permittivity of 4 leads
to robust SCF convergence of proteins in vacuum~\cite{Antony_2012}
imply that screening of the surface dipole is sufficient to restore
the HOMO-LUMO gap.
In the following, we test a number of methods for setting up a QM
cluster in such a way that DFT calculations employing either pure or hybrid
functionals may be readily applied.

The electric field across the cluster will be reduced if the atom
positions are allowed to relax, either by geometry optimisation or an
annealing procedure.
Figure~\ref{fig:gap+dip_plot}(b) (blue line) reveals that, following
classical minimisation, via fast conjugate gradient and Newton-Raphson
optimisation (Supplementary Methods), the average dipole moment of the
extracted water clusters is substantially reduced (as measured by
classical point charges).
Furthermore, the HOMO-LUMO gaps of clusters that have undergone MM
minimisation are all restored to values close to the bulk water value
of 4.2~eV (Figure~\ref{fig:gap+dip_plot}(a) (blue line)).
In addition, an implicit solvent model is expected to reduce the shift in electronic
states on opposite surfaces by screening the electrostatic potential
across the cluster.
Figure~\ref{fig:gap+dip_plot}(a) (red line) shows that when the
extracted water cluster is simulated with implicit solvent in ONETEP the HOMO-LUMO
gap is again restored to the bulk value.
Figures~\ref{fig:LDoS_plot_16}(b,c) confirm that the dipole moment of
the 16~\AA{} water cluster is reduced following MM minimisation and is
negligible in the dielectric medium.
In both cases, the density of electronic states is less dependent on
the z-coordinate and closely resembles the bulk DoS (green line), as
expected for a large cluster.
%
\section{Protein Systems}\label{sec:ProteinSystems}

\begin{figure}[h]
\centering
\includegraphics[scale=0.55]{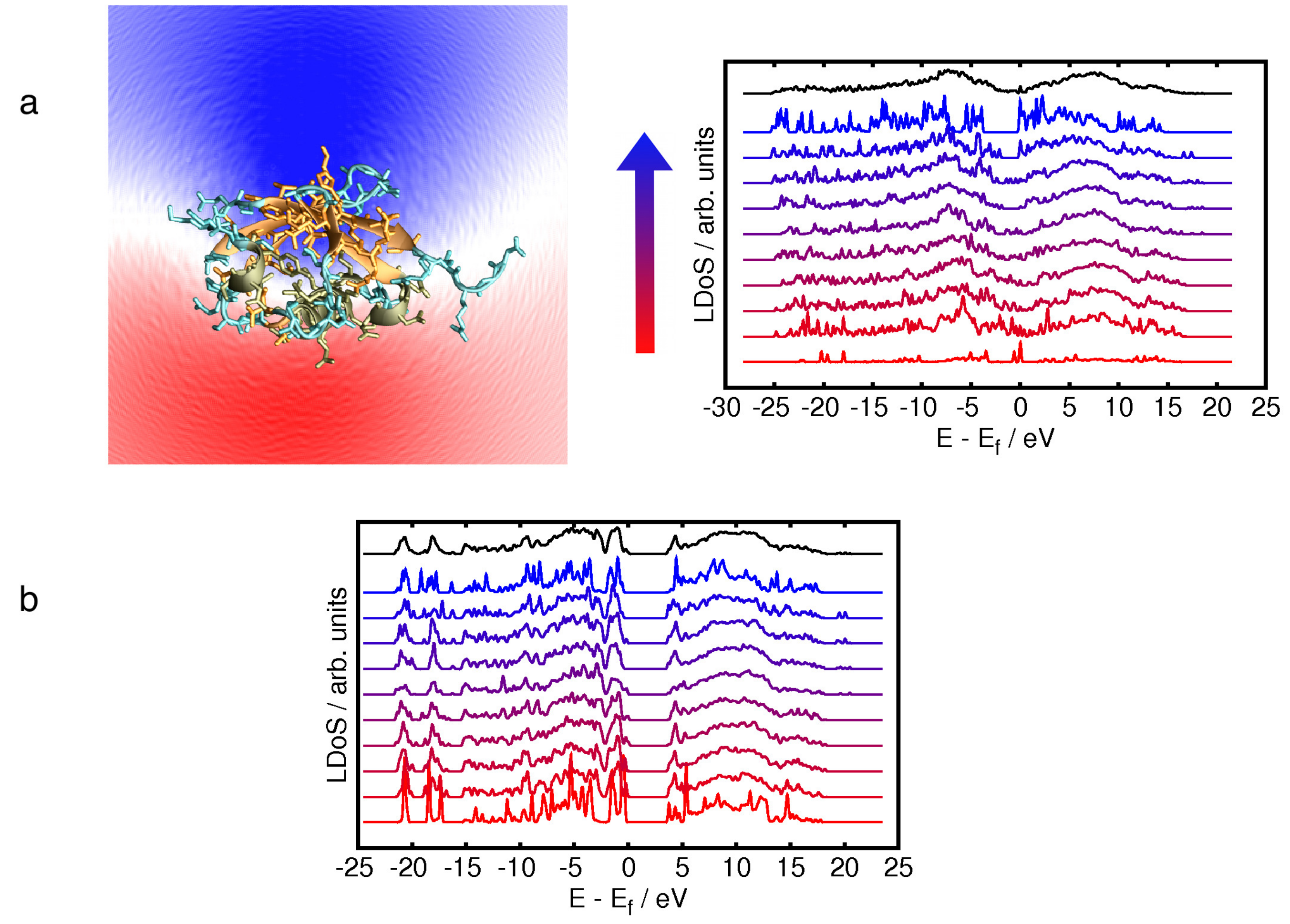}
\caption{Electrostatic potential and LDoS for groups of atoms as a
  function of position along the dipole moment vector (coloured arrow)
  of ubiquitin.  The
  black line is the total density of states.  Each line in the LDoS
  plot is normalised by the number of molecules contained in the slab.
  (a) The experimental structure with no solvation. The electrostatic
  potential ranges from -0.2 V (red) to +0.2 V (blue). 
  The slice is 42.9~\AA~behind the water cluster. (b) LDoS
  along the dipole moment of the same structure simulated with implicit solvent.}
\label{fig:LDoS_1UBQ}
\end{figure}

Turning our attention to polypeptide systems, six protein
conformations (1PLW~\cite{1PLW_citation}, 1FUL~\cite{1FUL_citation},
1RVS~\cite{1RVS_citation}, 1EDW~\cite{1EDW_citation},
1FDF~\cite{1FDF_citation} and 1UBQ~\cite{1UBQ_citation})
from the Brookhaven National Laboratory Protein Data Bank
(PDB)~\cite{PDB_citation} were used as starting configurations for our
calculations.
Table \ref{tab:Gap_values} shows the computed HOMO-LUMO gaps for these
structures \emph{in vacuo}, using the PBE functional.
In agreement with Ref.~\cite{Rudberg_2012} and as expected for this
method of system preparation, the calculations did not produce
HOMO-LUMO gaps (and hence did not converge) for any of the proteins
apart from the smallest system considered.
Figure \ref{fig:LDoS_1UBQ} reveals that the problem is similar in
nature to that of the water cluster.
Plotting the electrostatic potential far from the ubiquitin (1UBQ) protein
reveals a strong dipole moment.
The LDoS reveals the dipole moment 
has a significant effect on the valence states, 
shifting them to where the HOMO-LUMO gap should lie.  

In order to recover the expected HOMO-LUMO gaps, we use similar
techniques to those described in the previous section (Supplementary
Methods).
Classical minimisation was performed on the 1FDF structure \emph{in
  vacuo}, but again the electronic structure calculation failed to
converge.
Classical optimisation, though able to restore the HOMO-LUMO
gap in water clusters, is unsuitable for protein systems.
This may be understood from considering the reduced opportunity for
mobility in proteins, compared to water, as residues are fixed in
secondary structure conformations.
In addition, proteins may contain residues that are charged at
physiological pH, yet when these are simulated \emph{in vacuo} the charges are
unscreened, and thus have a stronger and longer-ranged effect on the
electrostatics than they do when solvated.
Such residues will contribute significantly to the molecular dipole
moment and may cause an undesired shift in the energies of the surface electronic states.
We hypothesise that a more suitable strategy is to include the effects
of the environment through the use of explicit water molecules or implicit solvent,
in order to screen the effect of charged residues.

To this end, each protein structure was solvated by a classically
minimised 5~\AA~layer of water and simulated in ONETEP using full DFT
for the entire system (up to 2386 atoms) to re-calculate the
electronic structure.
In addition, implicit solvent calculations have been used for the protein structures
in their vacuum configurations.
Using either an explicit or implicit solvation strategy restores the
HOMO-LUMO gaps to similar values.
Figure \ref{fig:LDoS_1UBQ}(b) shows that the DoS for the implicitly
solvated system more closely resembles that of an insulator.
The associated variation in the electrostatic potential is negligible, as in the case of the water cluster from 
Figure \ref{fig:LDoS_plot_16}(c) above, 
and is not shown.
To explore the possibility of reducing the computational cost of the
calculation, we have also represented the explicit water layer by
embedded point charges with a TIP3P charge distibution.
In this case, the HOMO-LUMO gap is very similar to that of the full QM
water layer, implying that classical charges produce an electrostatic environment 
appropriate to reduce the net dipole of the system.

\begin{table}[h]
		\begin{tabular}[h]{ccccccc}\cline{1-7}\cline{1-7}
\multicolumn{3}{c}{}&\multicolumn{4}{c}{HOMO-LUMO Gap / eV}	\\ \cline{4-7}				
\multicolumn{1}{c}{PDB ID}&Atoms& Charge &	\emph{in vacuo}	&	QM water  & Implicit Solvent & QM/EE	\\	\cline{1-7}
\multicolumn{1}{c}{1PLW}&	75(456)	&	0 & 0.0 & 3.7 & 3.7 & 3.5	\\	
\multicolumn{1}{c}{1FUL}&	135(453)	&	-1 & n/a & 2.7 & 2.6 & 2.6	\\
\multicolumn{1}{c}{1RVS}&	172(670)	&	0 & n/a & 3.4 & 3.7 & 2.9	\\ 
\multicolumn{1}{c}{1EDW}&	399(978)	&	-1 & n/a & 3.1 & 3.7 & 2.6\\
\multicolumn{1}{c}{1FDF}&	419(1526)	&	3 & n/a & 1.9 & 3.3 & 1.6 \\	   
\multicolumn{1}{c}{1UBQ}&	1231(2386)	&	0 & n/a & 2.6 &3.4& 2.4  \\\cline{1-7}\cline{1-7}
		\end{tabular}
		\caption{HOMO-LUMO gaps for a range of proteins from the PDB. Atom number in parentheses includes a 5 \AA~solvation shell of water used in classical minimisation and QM/EE simulations. Systems that did not converge are indicated by n/a. Vacuum calculations and implicit solvent simulations did not include any explicit water molecules.}\label{tab:Gap_values}
\end{table}
\section{Conclusions}\label{sec:Con}

We have confirmed recent findings~\cite{Rudberg_2012} that DFT
electronic structure optimisation is hindered by vanishing HOMO-LUMO
gaps in large water and protein clusters -- systems that should
display insulating behaviour.
This issue manifests itself in clusters prepared with improper
treatment of the vacuum/system interface.
In the examples presented here, unequilibrated vacuum/water interfaces
and X-ray protein crystal structures exhibit strong dipole moments.
Our LDoS calculations, decomposed by slabs along the direction of the
dipole moment, reveal that electronic states are shifted by the
electric field across the cluster.
The Fermi energy is, hence, pinned by states on opposite surfaces, 
which leads to closure of the HOMO-LUMO gap.
The presence of a 4.2 eV band gap in a 2010-atom model of bulk water calculated with the 
PBE exchange-correlation functional, emphasises that this effect has an electrostatic origin, 
rather than being peculiar to any particular class of functionals.
Hybrid functionals that tend to have an intrinsically wider gap appear
to be able to converge clusters comprising thousands of atoms \cite{Kulik_2012,Antony_2012}, 
but we would
expect HOMO-LUMO gap closure, even for functionals containing
Hartree-Fock exchange, once DFT methodological advances allow access
to still larger systems.

In this communication, we have demonstrated practical solutions for
restoring the HOMO-LUMO gap, with methods ranging from classical
structural optimisation of water/vacuum interfaces, to screening of
molecular dipole moments via implicit solvation of protein structures.
In general, implicit solvation gives the best correspondence between
HOMO-LUMO gaps of large water clusters and the bulk band gap, and
restores larger HOMO-LUMO gaps for proteins than does explicit
solvation.
In the present work we have looked at systems comprising up to 2386
atoms, and the practical solutions demonstrated here will allow the
continued investigation of biomolecular systems through the use of
Kohn-Sham DFT.

\ack
The authors acknowledge Chris-Kriton Skylaris for useful discussions and Louis Lee for the use of 1UBQ calculations.
Computational resources were provided by the Cambridge HPC Service, funded by EPSRC Grant EP/F032773/1. PDH acknowledges the support of a Royal 
Society University Research Fellowship. NDMH acknowledges 
the support of EPSRC grant EP/G05567X/1 and a Leverhulme Early Career Fellowship. DJC and GL acknowledge support from the EPSRC. 

\section*{References}
\bibliographystyle{unsrt}
\bibliography{gap_v1.11}

\section*{Supplementary Methods}

\subsection*{AMBER parameters}

 Water molecules were generated using the tleap module of AMBER with associated TIP3P charges. 
 Coulomb interactions were treated using the 
 Particle Mesh Ewald sum, with a real space cutoff of 10~\AA. 
 The cut-off length for Lennard-Jones interactions was also 
 set to 10~\AA. 
 For the 4763 atom water periodic cube, the system was minimised in the NVT ensemble before being heated to 300K 
 in six stages in the NPT  ensemble. 
 A production run of 5 ns at 300 K was then performed and snapshots were saved every 1 ps. 
The 2010 atom structure for the bulk water HOMO-LUMO gap calculation in ONETEP underwent identical preparation. 

With regards to the protein systems, in the event that starting configurations obtained from the PDB had more than one conformation 
available, the structure labelled as `model 1' was used.
 All protein interactions 
 were described using the AMBER ff99SB biomolecular force field~\cite{Hornak_06}. 
The protein systems were solvated in a 50~\AA~water cube with a TIP3P
charge distribution.
NVT minimisation was performed before NPT equilibrating to 300K in six steps. 
A 5 ns NVT production run 
was then performed to generate the final structures. 
Throughout minimisation, equilibration and the production run, harmonic constraints of 100 kcal mol$^{-1}$~\AA$^{-2}$  
were imposed upon the protein.
The water molecules were then stripped from the structure and these
protein configurations were used as our vacuum conformations.
In order to recover the expected HOMO-LUMO gaps, we use similar
techniques to those described in the main text.  
Namely, 5~\AA~of the surrounding water molecules were retained from
the molecular mechanics equilibration simulations for each protein and the solvent
geometry was optimised via fast conjugate gradient (CG) followed by Newton-Raphson (NR) minimisation until the
root mean square force decreased below 10$^{-4}$ kcal mol$^{-1}$~\AA$^{-1}$ during CG minimisation and 
below 10$^{-10}$ kcal mol$^{-1}$~\AA$^{-1}$ for NR minimisation, whilst
the protein residues remained fixed.

\subsection*{ONETEP parameters}

Interactions between electrons and nuclei were described by norm-conserving 
pseudopotentials. 
NGWFs were initialized as orbitals obtained from 
solving the Kohn-Sham equation for free atoms, with a $1s$ configuration for H, a $2s2p$ configuration 
for C, N, O and a $3s3p$ configuration for S~\cite{onetep_orbitals}. 
The NGWFs were expanded in a basis 
of periodic cardinal sine (psinc) functions~\cite{onetep_4} with an energy cut-off of 916 eV, corresponding to 
a grid spacing of 0.475 a$_0$, and were localised in real space 
with radii of 5.3~\AA. 
In the case of the 1PLW protein, an increase in NGWF radii from 
5.3~\AA~to 6.4~\AA~led to the change of the HOMO-LUMO gap by 3 meV. 
These radii at a cut-off of 1020 eV give a change in HOMO-LUMO gap of 6 meV when compared to a 916 eV cut-off energy calculation. 
NGWFs are optimised \emph{in situ} to represent the valence states: however,
previous experience shows these describe the conduction states well for
at least the first 1-2~eV above the LUMO and thus produce the same gap as
equivalent plane-wave calculations. 
At energies beyond this point,
however, the DoS is not well-represented by NGWFs \cite{Ratcliff_2011} and
should be discounted.
The spherical cut-off approach for Coulomb potentials was 
used to eliminate all interactions of the molecules with their periodic images~\cite{onetep_coulomb}. 

Where indicated in the text, implicit solvent has been used.
Using a smeared-ion formalism, the molecular Hartree energy is obtained not in reciprocal space, like standard ONETEP, but rather by solving the Poisson equation (homogeneous in vacuum, inhomogeneous in solution) in real space, under open boundary conditions.
This is achieved via a multigrid approach detailed elsewhere \cite{Dziedzic_2011, Dziedzic_2012}.
A 10 $a_0$ gap is left between the electron density and the edge of the multigrid and the ion smearing width is 0.8~a$_0$.
The converged electronic density obtained in vacuum was used to generate the density-dependent dielectric cavity in solution.
The values of the solvation parameter $\beta$ and electronic density threshold $\rho_0$ were 1.3 and 3.5$\times 10^{-4}$ a.u. respectively, as proposed in \cite{Scherlis}.
The relative dielectric permittivity of the solvent was set to 80.0 for all implicit solvent calculations.

\newpage \section*{Dipole moment as a function of system size}

\begin{figure}[h]
\centering
\includegraphics[scale=0.35]{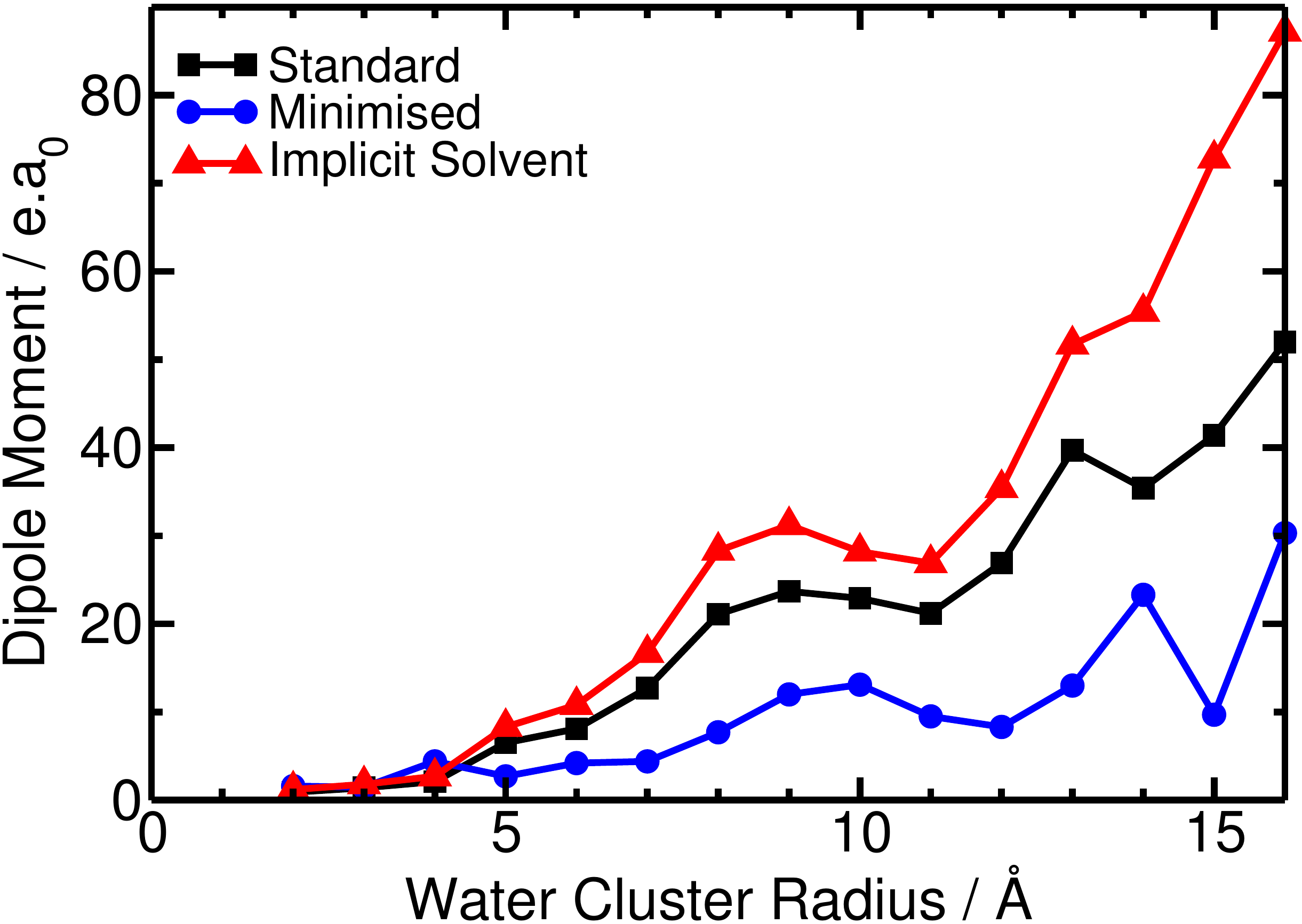}
\caption{Calculated total dipole moment of water clusters of
increasing radius. Black Line: ONETEP calculated QM dipole moment from the
final molecular dynamics snapshot at each radius. The dipole moment increases with radius in 
the same manner as in our classical simulations. Blue line: ONETEP calculated
dipole on the same snapshot after classical minimisation, which reduces the dipole moment across the cluster. 
Red line: ONETEP implicit solvent calculations. In this case, the dielectric medium supports a higher dipole moment, 
although the net potential is screened at large distances.}
\label{fig:SI_dip}
\end{figure}

\end{document}